\documentclass[iop]{emulateapj}
\usepackage{epstopdf} 

\newcommand{\ngoldstars}{\ensuremath{518}}
\newcommand{\ngoldspectra}{\ensuremath{834}}
\newcommand{\ngoldstarsdwarfs}{\ensuremath{88}}
\newcommand{\ngoldstarsdwarfslogg}{\ensuremath{4.2}}
\newcommand{\nSTP}{\ensuremath{157}}
\newcommand{\nSTPunique}{\ensuremath{119}}

\newcommand{\rpLimit}{\ensuremath{1.7~\rearth}}

\newcommand{\kms}{\ensuremath{\mathrm{km\,s^{-1}}}}

\newcommand{\angstrom}{\ensuremath{\rm{\AA}}}

\newcommand{\dex}{\ensuremath{\rm{dex}}}

\newcommand{\rearth}{\ensuremath{R_\earth}}
\newcommand{\mearth}{\ensuremath{M_\earth}}

\newcommand{\rpl}{\ensuremath{R_{p}}}
\newcommand{\mpl}{\ensuremath{M_{p}}}

\shorttitle{Metallicity of nonTEP}
\shortauthors{Buchhave et al.}

\begin{document}

\title{The metallicities of stars with and without transiting planets}

\altaffiltext{1}{Harvard-Smithsonian Center for Astrophysics, Cambridge, MA 02138, USA}
\altaffiltext{2}{Centre for Star and Planet Formation, Natural History Museum of Denmark, University of Copenhagen, DK-1350 Copenhagen, Denmark}

\author{
	Lars~A.~Buchhave\altaffilmark{1,2}
	David~W.~Latham\altaffilmark{1}
}

\begin{abstract}

Host star metallicities have been used to infer observational constraints on planet formation throughout the history of the exoplanet field. The giant planet metallicity correlation has now been widely accepted, but questions remain as to whether the metallicity correlation extends to the small terrestrial-sized planets. Here, we report metallicities for a sample of \ngoldstars\ stars in the Kepler field that have no detected transiting planets and compare their metallicity distribution to a sample of stars that hosts small planets ($\rpl < 1.7~\rearth$). Importantly, both samples have been analyzed in a homogeneous manner using the same set of tools (Stellar Parameters Classification tool; SPC). We find the average metallicity of the sample of stars without detected transiting planets to be $\rm{[m/H]}_{SNTP,dwarf} = -0.02 \pm 0.02~\dex$ and the sample of stars hosting small planets to be $\rm{[m/H]}_{STP} = -0.02 \pm 0.02~\dex$. The average metallicities of the two samples are indistinguishable within the uncertainties, and the two-sample Kolmogorov-Smirnov test yields a p-value of 0.68 ($ 0.41~\sigma$), indicating a failure to reject the null hypothesis that the two samples are drawn from the same parent population. We conclude that the homogeneous analysis of the data presented here support the hypothesis that stars hosting small planets have a metallicity similar to stars with no known transiting planets in the same area of the sky.

\end{abstract}
\keywords{	planetary systems ---
	techniques: spectroscopic, spectroscopic – surveys}

\section{Introduction}
\label{sec:intro}

The discovery of thousands of transiting exoplanets has enabled studying the properties of statistically significant ensembles of planets and their host stars. Such studies, accompanied by planet formation theories, can be used to shed light on how planets form. In early stages of the exoplanet field when only a handful of planets were known to exist, a correlation between host star metallicity and the presence of hot-Jupiter type planets was suggested \citep{gonzalez_stellar_1997}. This tendency was later confirmed by a series of papers showing that metal-rich stars are much more likely to harbor gas-giant planets \citep{santos_spectroscopic_2004,fischer_planet-metallicity_2005}. This result is usually interpreted as natural support for the core accretion model, where planets growing in a metal-rich environment will have a higher likelihood of reaching the critical core mass allowing run-away gas accretion before the gas in the system has dissipated \citep{ida_toward_2005,mordasini_extrasolar_2009,mordasini_extrasolar_2012}. Recent discoveries of thousands of small exoplanets have allowed these types of studies to extend into the terrestrial planets regime. 

The hot-Jupiter type exoplanets with periods shorter than about 10 days dominated the initial discoveries, but we now know these types of planets to be rare \citep[found around $\sim 1 \%$ of Sun-like stars;][]{wright_frequency_2012}. On the other hand, the Kepler Mission has demonstrated that small planets are astonishingly common and ubiquitous in our galaxy \citep{howard_planet_2012,fressin_false_2013,petigura_prevalence_2013,dressing_occurrence_2013} thus supporting and extending the earlier hints from radial velocity surveys that low-mass planets are common \citep{mayor_quest_2008,howard_nasa-uc_2009,bonfils_harps_2013}. Initially, the metallicity correlation was assumed to pertain to exoplanets in general. However, a number of later studies have shown that the giant-planet metallicity correlation does not seem to extend into the small planet regime \citep{sousa_spectroscopic_2011,buchhave_abundance_2012,neves_metallicity_2013,everett_spectroscopy_2013}. Recent results measuring the metallicities of a large number of stars hosting small planets suggest that the planets can be divided into three regimes, namely terrestrial planets, gas-dwarf planets (small planets with lower mean densities) and gas-giants \citep{buchhave_three_2014}, where the planets in the terrestrial planet regime have a metallicity consistent with solar. Then, on the other hand, \cite{wang_revealing_2015} used photometrically derived metallicities from the Kepler Input Catalog \citep{brown_kepler_2011} combined with spectroscopic  metallicities from \cite{buchhave_three_2014} to suggest that the occurrence rate of small planets is higher around metal rich stars, thus claiming that the planet-metallicity correlation does extend all the way down to the small planets. 

The giant planet-metallicity correlation was determined by examining the metallicities of a sample of host stars harboring hot-Jupiters and comparing it to a ``control sample'' of stars that did not host such planets. This made sense, since the Doppler surveys could readily detect these massive close-in companions except for the rare cases where the orbital plane of the system was nearly face-on. A control sample of stars without hot-Jupiter companions was thus easy to construct for a volume limited survey. As the radial velocity surveys increased their sensitivity and were able to detect sub-Neptune sized planets, it was still possible to construct a sample of stars that did not host planets of the detected type and estimate how the planet occurrence rate was affected by host star metallicity \citep[e.g.][]{sousa_spectroscopic_2011}. 

However, it is much more difficult to construct a control sample for the small transiting exoplanets discovered by Kepler. We know that small planets are common around solar-type stars, and while the transit method has the required sensitivity to detect such small planets, the orbital plane must be oriented such that the planets transit for us to detect them. As a result, we cannot construct a sample of stars free of planets from using transit surveys. The current level of precision of the leading radial velocity spectrographs is not yet able to detect the small masses of these planets, except in rare circumstances. In fact, it turns out to be much more difficult to construct a sample of stars where we can definitively rule out the presence of small planets than to discover the planets themselves.

In this paper, we examine a sample of stars in the Kepler field with no detected transiting planets (Stars with No detected Transiting Planets: SNTP) and compare their metallicities to stars hosting planets with sizes below  $\rpl < 1.7~\rearth$ (Stars with detected Transiting Planets: STP).

\section{Samples and observations}
\label{sec:sampleandobs}

Our sample of SNTP consists of \ngoldstars~stars previously analyzed with asteroseismology \citep{chaplin_asteroseismic_2014}. The stars were selected for observation by the Kepler Follow-up Program team (KFOP) as a sample of ``gold standard'' stars that would be well-characterized by both asteroseismology and spectroscopy. The stars were observed with the Tillinghast Reflector Echelle Spectrograph (TRES) on the 1.5 m Tillinghast Reflector at the Fred Lawrence Whipple Observatory on Mt. Hopkins, Arizona using the medium resolution fiber ($2\farcs3$ projected diameter) with a resolving power of $R \simeq 44\,000$, giving a wavelength coverage of $\sim 3800-9100~\angstrom$.

Most of the stars were observed only once, but in some cases, several observations were available. In total, we analyzed \ngoldspectra\ spectra of \ngoldstars\ stars. Since the magnitude of the host stars ranged from 3rd to 13th magnitude, the exposure time ranged from 45 seconds to 1 hour with a mean exposure time of approximately 11 minutes. The resulting average signal-to-noise ratio per resolution element (SNRe) in the Mg b region was 49 corresponding approximately to a signal-to-noise ratio per pixel (SNR) of 30. The spectra used in this paper are available on the Kepler Community Follow-up Observing Program (CFOP) website (http://cfop.ipac.caltech.edu).

\section{Stellar parameters}
\label{sec:stellarparams}

\begin{figure}
	\centering
	\epsscale{1.2}
	\plotone{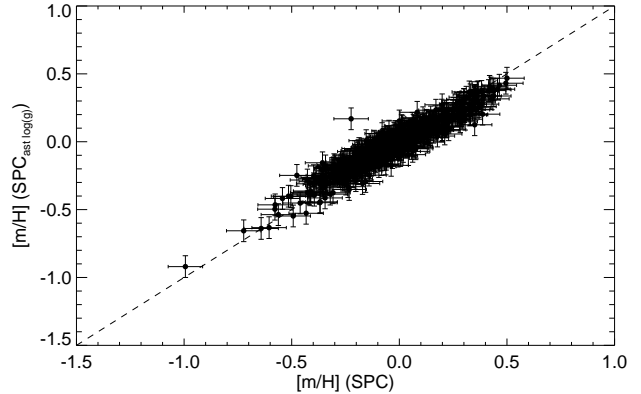}
	\caption{The metallicities of the stars in the SNTP sample determined using SPC with all parameters free versus SPC using the surface gravity from asteroseismology as a prior. The mean difference is -0.01 dex and the RMS is 0.07 dex, suggesting that metallicities can be determined reliably without a prior on surface gravity.}
	\label{fig:mh_astero}
\end{figure}

The sample of SNTP stars was analyzed with the Stellar Parameters Classification tool (SPC) in the same manner as documented in \cite{buchhave_abundance_2012} and \cite{buchhave_three_2014}. Here, we provide a brief overview of SPC.

SPC cross-correlates the observed spectrum against a grid of synthetic template model spectra with varying values of stellar parameters for effective temperature, surface gravity, metallicity and rotational velocity. The normalized cross-correlation function (CCF) peak height is determined for each template, indicating how well the observed spectrum matches the various synthetic model spectra. The library consists of an extensive grid of stellar models \citep{kurucz_model_1992} and covers the wavelength region between $5050 - 5360~\angstrom$. It spans an interval for effective temperature of $3500~\rm{K} < T_{eff} < 9750~\rm{K}$, surface gravity of $0.0 < \rm{log}(g) < 5.0$, metallicity of $-2.5 < \rm{[m/H]} < +0.5$ and rotational velocity of $0~\kms < V_{\rm{rot}} < 200~\kms$ and has a spacing of 250~K in effective temperature, 0.5 in $\rm{log}(g)$, 0.5~dex in $\rm{[m/H]}$ and progressive spacing in rotational velocity yielding a grid consisting of 51,359 synthetic spectra. 

Rather than selecting the best matched template, which would restrict the stellar parameters to the rather coarse grid spacing, SPC assumes that the normalized CCF peak height varies smoothly between grid points. The CCF peaks are fitted with a three dimensional third order polynomial as a function of effective temperature, surface gravity and metallicity, weighting the peaks proportional to their height, giving less weight to the CCF peaks that are a poor match to the spectrum. The peak of the fitted surface is  determined, yielding the final stellar parameters.

SPC utilizes the entire spectrum in the wavelength region of the library, enabling the technique to derive stellar parameters from spectra with relatively low SNR (down to a SNRe of roughly 30 corresponding to a SNR of 18 per pixel). Since SPC takes advantage of all the absorption lines in the wavelength region, we denote the metallicities in this work by [m/H], representing a mix of metals assumed to be the same as the relative pattern of the abundances in the Sun, not to be confused with the abundances for individual elements (e.g. [Fe/H]).

SPC determines all the stellar parameters simultaneously, however, if some parameters are available from a more reliable source, we can use this information to set a prior on these parameters in SPC. The surface gravity is notoriously difficult to determine spectroscopically \citep[e.g.][]{torres_improved_2012}, so we used the value determined from asteroseismology in \cite{chaplin_asteroseismic_2014} as a prior. Since the sample of STP stars does not have surface gravities determined by asteroseismology, we investigated whether enforcing a prior on the surface gravity would significantly affect the metallicities by running SPC with all parameters allowed to float and using the asteroseismic surface gravities as a prior. The resulting metallicities are shown in Figure \ref{fig:mh_astero}. The mean metallicity difference between the two sets of metallicities is -0.01 dex and the RMS is 0.07 dex, indicating that using a prior on the surface gravity has little effect on the metallicities. The metallicities of the SNTP stars are listed in Table \ref{tab:metallicities}.

In order to determine the mass and radius of a star, asteroseismology relies on effective temperatures and metallicities from e.g. spectroscopy or photometry. In \cite{chaplin_asteroseismic_2014}, photometrically derived effective temperatures were available for the entire sample, but since no reliable metallicities were available, the metallicities were set to $-0.2 \pm 0.3~\dex$. Spectroscopic estimates of effective temperature and metallicity were available for a subset of 87 stars. In this paper, we focus on the metallicities of the host stars, but a second iteration of the asteroseismic results using stellar parameters from SPC with a asteroseismic prior on the surface gravity is the subject of a paper in preparation where all the stellar parameters will be published. However, experience shows that this iteration only has a minor effect on the final stellar parameters and in particular the metallicity, as demonstrated in Figure \ref{fig:mh_astero}.

\begin{deluxetable}{lr}
	\tablewidth{7cm}
	\tablecaption{
		Metallicities of star with no detected transiting planets
		\label{tab:metallicities}
	}
	\tablehead{
		\colhead{KIC} &
		\colhead{[m/H]} 
	}
	\startdata
1430163 & $-0.05 \pm 0.08 $ \\
1435467 & $ 0.01 \pm 0.08 $ \\
1725815 & $-0.07 \pm 0.08 $ \\
2010607 & $-0.01 \pm 0.08 $ \\
2306756 & $ 0.42 \pm 0.08 $ \\
....... &  \\
[-1.5ex]
	\enddata
\tablecomments{Table \ref{tab:metallicities} is published in its entirety in the electronic edition of the Astrophysical Journal.}	
\end{deluxetable}

\section{Metallicities of stars with and without detected transiting planets}
\label{sec:metallicities}

\begin{figure}
	\centering
	\epsscale{1.2}
	\plotone{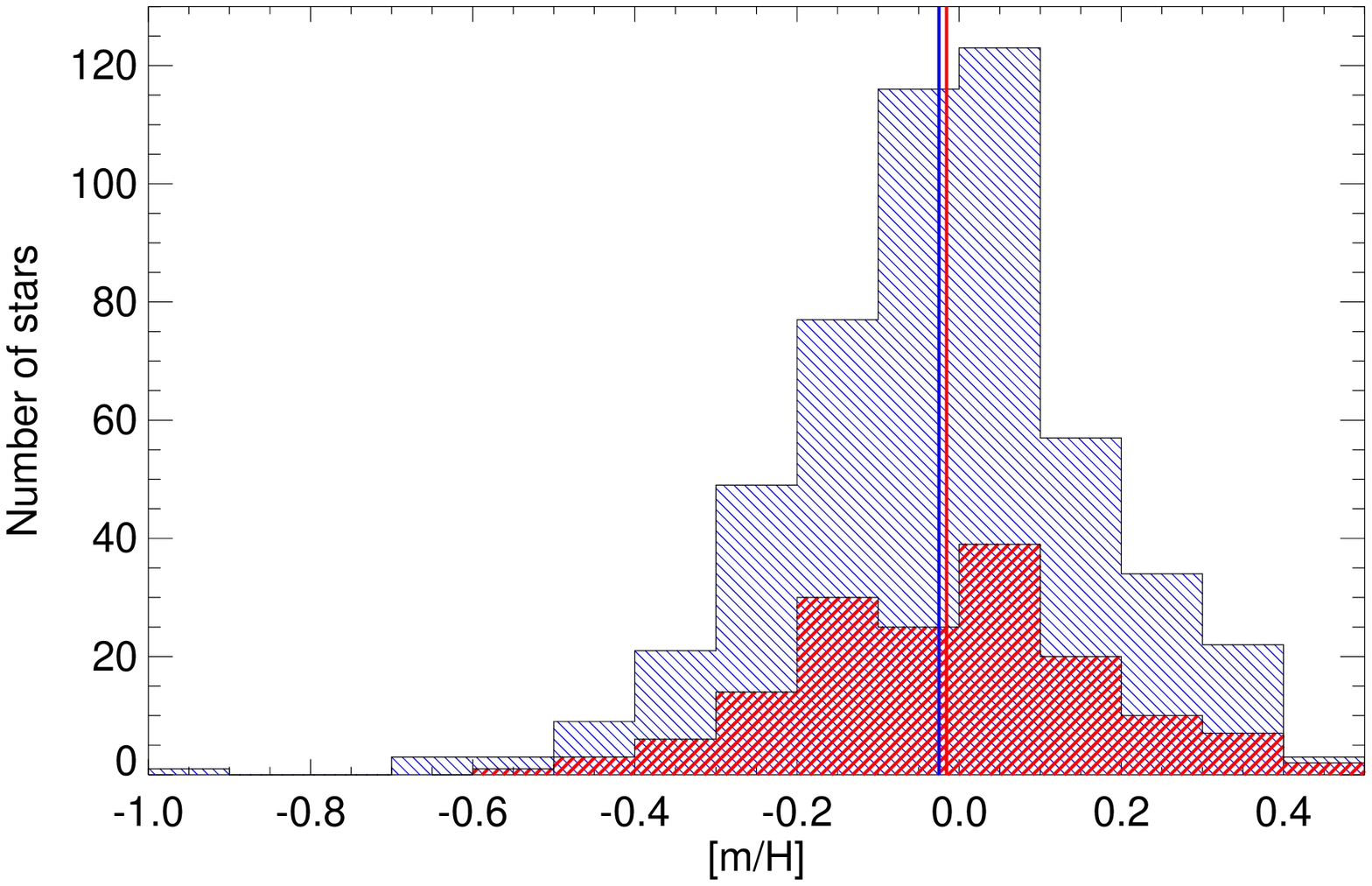}
	\caption{The distribution of metallicities of the stars with no detected transiting planets (blue, \ngoldstars~stars) and the sample of stars with small planets (red, \nSTP~stars) from \cite{buchhave_three_2014}. The vertical lines show the average metallicity of the two samples: $\rm{[m/H]}_{SNTP} = -0.02 \pm 0.01~\dex$ and $\rm{[m/H]}_{STP} = -0.02 \pm 0.02~\dex$, where the uncertainty is the standard error of the samples.}
	\label{fig:mh_all}
\end{figure}
\begin{figure}
	\centering
	\epsscale{1.2}
	\plotone{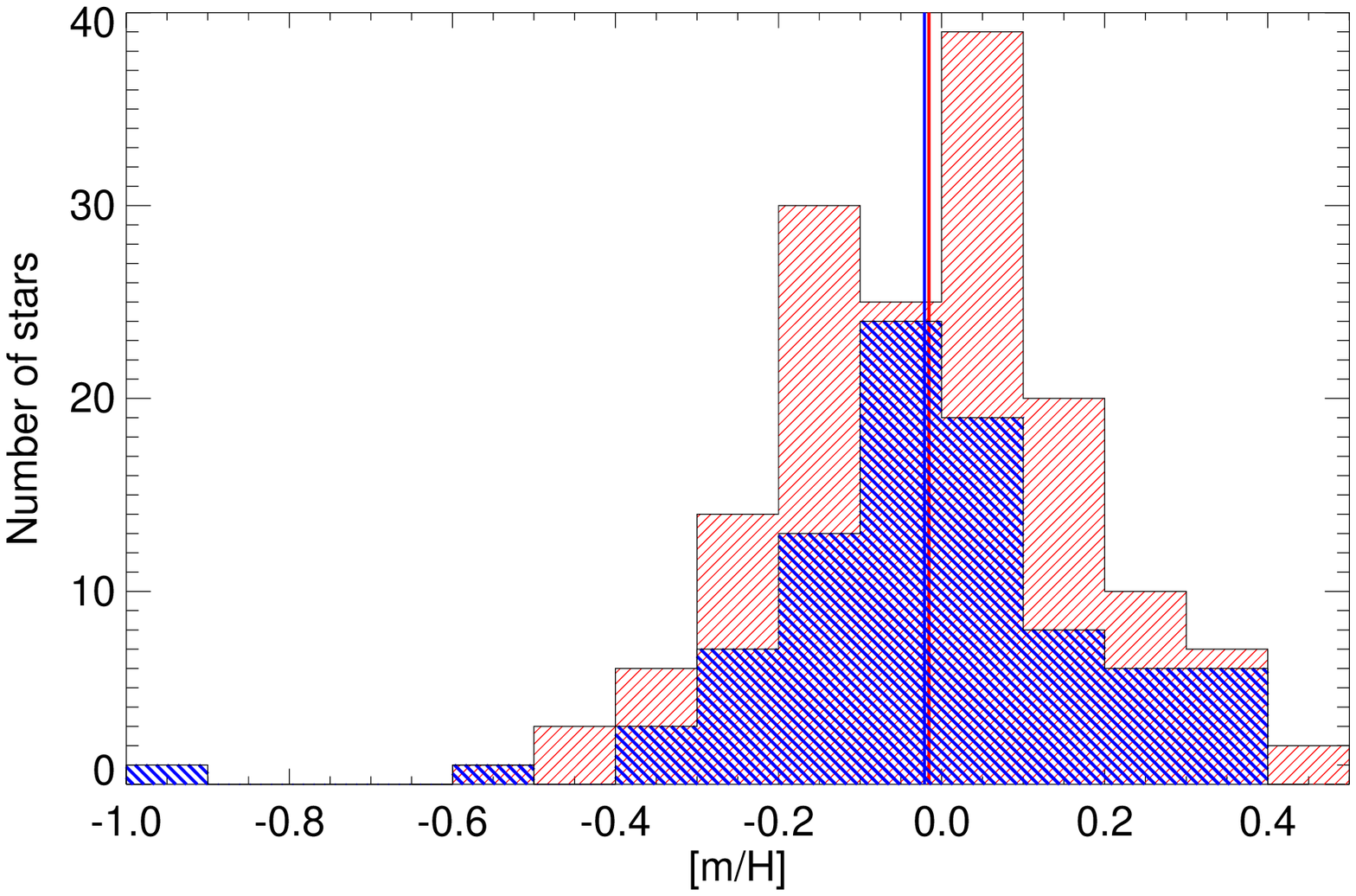}
	\caption{The distribution of metallicities of dwarf stars with no detected transiting planets  (blue, $\rm{log}(g) > \ngoldstarsdwarfslogg$, \ngoldstarsdwarfs~stars) and the sample of stars with small planets (red,  \nSTP~stars) from \cite{buchhave_three_2014}. The vertical lines show the average metallicity of the two samples: $\rm{[m/H]}_{SNTP,dwarf} = -0.02 \pm 0.02~\dex$ and $\rm{[m/H]}_{STP} = -0.02 \pm 0.02~\dex$, where the uncertainty is the standard error of the samples.}
	\label{fig:mh_dwarf}
\end{figure}

We have determined the metallicities of a sample of stars in the Kepler field that have no detected transiting planets (SNTP). Although the transit method has not detected any planets, many of the stars in the sample are most likely planetary hosts, since the occurrence rates of small planets is high \citep{howard_planet_2012, fressin_false_2013, petigura_prevalence_2013}. Unfortunately, there is currently no feasible way to construct a ``control sample'' of stars free of small planetary companions, since our current methods are either not sensitive enough or cannot exclude the presence of planets. Nevertheless, if some of the stars in the SNTP sample do not host planets and if such stars are on average more metal poor then planet hosting stars, we would expect the average metallicity of the SNTP sample to be lower than that of a sample of stars hosting planets.


We compare the metallicities of the SNTP stars to samples of stars hosting transiting planets  (STP) smaller than $\rpl < 1.7~\rearth$ from \cite{buchhave_three_2014}. Short period planets receive a large amount of a incident flux from their host star and could as a result undergo significant evaporation of their atmospheres, which is especially true for smaller planets \citep{owen_kepler_2013}. Such planets could have accreted a gaseous atmosphere when they formed and would later have this atmosphere stripped off by photo evaporation. This will decrease the planet radii and could therefore potentially distort the metallicity-radius correlation we are attempting to quantify. Following \cite{buchhave_three_2014}, we therefore remove highly irradiated planets ($F_\nu > 5 \times 10^5~\rm{J~s^{-1}~m^{-2}}$) from the sample.

Figure \ref{fig:mh_all} shows the distribution of metallicities of the two samples (stars with and without detected transiting planets). The SNTP sample contains \ngoldstars~ stars and the STP sample contains \nSTP~stars. The average metallicity of the two samples is $\rm{[m/H]}_{SNTP} = -0.02 \pm 0.01~\dex$ and $\rm{[m/H]}_{STP} = -0.02 \pm 0.02~\dex$, where the uncertainty is expressed as the standard error of the mean. As mentioned, the stars in the SNTP sample were selected because they have asteroseismic detections. Since asteroseismic oscillations are easier to detect in evolved stars, the SNTP sample is dominated by sub-giants, while the STP sample primarily contains dwarf stars. To check whether the dominating sub-giants affect the average metallicity, we exclude stars with weak gravities ($\log(g) < \ngoldstarsdwarfslogg$) from the SNTP sample. Figure \ref{fig:mh_dwarf} shows the distribution of the metallicities of the $\rm{SNTP_{dwarf}}$ stars (\ngoldstarsdwarfs~stars) compared with the sample of STP stars. The average metallicity of the SNTP sample containing only dwarfs is essentially unaffected: $\rm{[m/H]}_{SNTP, dwarf} = -0.02 \pm 0.02~\dex$.

We performed a two-sample Kolmogorov-Smirnov (K-S) test to determine whether the metallicities of the two distributions of host stars are not drawn randomly from the same parent population. We find a p-value of 0.50 ($ 0.67~\sigma$) when using all the stars in the SNTP sample and a p-value of 0.68 ($ 0.41~\sigma$) when only including the dwarfs. This suggests that the K-S test fails to reject the null hypothesis that the two samples are drawn from the same parent population.

We also compare the metallicities of stars hosting larger ``gas-dwarf'' planets $1.7~\rearth < \rpl < 4.0~\rearth$ ($\rm{STP_{gd}}$) from \cite{buchhave_three_2014} to the stars with no detected transiting planet sample ($\rm{SNTP_{dwarf}}$). The average metallicity of the $\rm{STP_{gd}}$ sample is 
$\rm{[m/H]}_{STP_{gd}} = 0.05 \pm 0.01~\dex$. A K-S test of the $\rm{STP_{gd}}$ and $\rm{SNTP_{dwarf}}$ samples yields a p-value of 0.0026 indicating that the two samples are not drawn from the same parent population at a $3.01~\sigma$ confidence level, in contrast to the STP sample of smaller planets with an average metallicity of $\rm{[m/H]}_{STP} = -0.02 \pm 0.02~\dex$ (in agreement with the findings of \cite{buchhave_three_2014}).

The sample of stars with transiting planets is constructed by adding the host star of each individual planet to the sample. For multi-planet systems, this can result in a star being added multiple times if it hosts several small planets. If we instead construct a sample of unique host stars, with no regard to the number of small planets they host, the number of STP stars reduces to \nSTPunique~stars. The average metallicity of the $\rm{STP_{uniq}}$ sample is  $\rm{[m/H]}_{STP,uniq} = +0.01 \pm 0.02$ and the K-S p-value is 0.40 ($ 0.84~\sigma$) when comparing to the $\rm{SNTP_{dwarf}}$ sample. Like the other results, the average metallicities of the two samples are consistent within the uncertainties and the p-value suggests that the K-S test fails to reject the null hypothesis that the two sample are drawn from the same parent population of stars.

\section{Summary and Discussion}
\label{sec:discussion}
We have compared the metallicities of a sample of stars with no detected transiting planets (SNTP) to the metallicities of a sample of stars hosting small planets (STP with $\rpl < \rpLimit$) from \cite{buchhave_three_2014}. If some of the stars in the SNTP sample do not host any planets and if these stars are, in fact, more metal poor as a universal metallicity correlation would suggest, we would expect the average metallicity of the sample of SNTP stars to be more metal poor than a sample of stars hosting small transiting exoplanets. We find the average metallicity of the two samples to be $\rm{[m/H]}_{SNTP, dwarf} = -0.02 \pm 0.02~\dex$ and $\rm{[m/H]}_{STP} = -0.02 \pm 0.02~\dex$.

These results seem at odds with the conclusions from \cite{wang_revealing_2015}, who suggest the existence of a universal planet-metallicity correlation extending all the way down to the terrestrial planets. \cite{wang_revealing_2015} find a p-value of 0.005 when comparing the metallicities of the sample of stars hosting small planets from \cite{buchhave_three_2014} to the large sample of Kepler Input Catalog stars \citep[KIC;][]{brown_kepler_2011} with no transiting planet detections, suggesting that the stars in the two samples are not drawn from the same parent population at a $2.8~\sigma$ confidence level. The p-value is vastly different from our results (p-value of 0.68), which suggests that the samples are, in fact, drawn from the same parent population.

The discrepancy could possibly be due to the fact that \cite{wang_revealing_2015} compare metallicities derived using two different techniques. \cite{brown_kepler_2011} state that the KIC metallicities are 0.17 dex more metal poor than spectroscopically derived metallicities using SME \citep{valenti_spectroscopy_1996}, which is attributed to the Basian prior used to derive the KIC metallicities being rather narrowly peaked around -0.10 dex. Although \cite{wang_revealing_2015} did attempt to correct for this offset, it can be difficult to compare metallicities derived using different methods, especially when looking for subtle differences in metallicities. We also note that \cite{wang_revealing_2015} did not remove the highly irradiated planets that may not obey the radius-metallicity correlation we are studying (see Section \ref{sec:metallicities}).

We find the average metallicities of the SNTP stars and the STP stars are consistent within the uncertainties. Both samples comprise a large number of stars that were analyzed in a homogeneous manner using the same tools, thus avoiding a potential bias between the samples. A two-sided K-S test yields a p-value of 0.68 ($ 0.41~\sigma$), suggesting a failure to reject the null hypothesis that the two samples are drawn from the same parent population. Other previous publications have reached similar conclusions \citep{sousa_spectroscopic_2011,buchhave_abundance_2012,neves_metallicity_2013,everett_spectroscopy_2013}, although the publications based on planets detected via radial velocities contain only few planets with masses below $\mpl < 10~\mearth$. We conclude that there is no evidence for an enhanced metallicity of Kepler stars hosting small planets ($\rpl < \rpLimit$) when compared to dwarf stars in the Kepler field with asteroseismic detections but no detected transiting planets.

\end{document}